\documentclass[aps,prc,onecolumn,notitlepage,superscriptaddress]{revtex4-1}
\usepackage{blindtext}
\usepackage{xcolor}
\usepackage{hyperref}
\usepackage{epsfig}
\usepackage{subfigure}
\usepackage{graphicx}
\usepackage[toc,page]{appendix}
\usepackage{epstopdf}
\usepackage{amsmath,amssymb,amsfonts}
\usepackage{array}
\usepackage{url}
\usepackage{hyperref}
\usepackage{lineno}
\usepackage{xspace}
\usepackage{listings}
\usepackage{float}
\usepackage{subfigure}
\usepackage{subfiles}
\usepackage{siunitx}

\def\tsc#1{\csdef{#1}{\textsc{\lowercase{#1}}\xspace}}
\tsc{WGM}
\tsc{QE}

\def \auau  {Au+Au}
\def \zrzr  {Zr+Zr}
\def \ruru  {Ru+Ru}
\def \pp    {\textit{p}+\textit{p}}
\newcommand{\sqrtsNN}{\mbox{$\sqrt{\mathrm{s}_{_{\mathrm{NN}}}}$}}

\begin{document}
%\linenumbers
%\let\WriteBookmarks\relax
%\def\floatpagepagefraction{1}
%\def\textpagefraction{.001}

% Main title of the paper
\title {Charged-Particle Multiplicity Dependence of Net-Proton Cumulants in  \zrzr~and~\ruru~Collisions at \texorpdfstring{\sqrtsNN}{}~=~\SI{200}{GeV}}

\author{The STAR Collaboration}

\begin{abstract}
% %%%%%======================
We present measurements of cumulants of event-by-event net-proton 
distribution at mid-rapidity and their ratios up to the sixth order as a 
function of charged-particle multiplicity in 
$^{96}_{40}$Zr+$^{96}_{40}$Zr and $^{96}_{44}$Ru+$^{96}_{44}$Ru
(isobars)~collisions at a nucleon–nucleon center-of-mass energy (\sqrtsNN) of 200 GeV. The data are collected from the STAR experiment with a total
of two billion events recorded for each collision system. 
The measurements are compared with those obtained from
\pp~and~\auau~collision systems at the same center-of-mass energy. 
The higher-order cumulant ratios ($C_4/C_2$, $C_5/C_1$, and $C_6/C_2$) show an overall decreasing trend as a function of the charged-particle multiplicity across systems. The isobar results align with the Au+Au trends within uncertainties. The observations are compared with calculations from Lattice Gauge Theory (LGT) that include a quark-hadron crossover. The systematic behavior of the higher-order cumulant ratios shows that, overall, they progressively approach LGT predictions with increasing multiplicity within uncertainties. This could imply that the medium created in these heavy-ion collisions gradually evolves into thermalized QCD matter undergoing a crossover transition with multiplicity.
\end{abstract}

\maketitle

%%%%%%%%%%%%%%%%%%%%%%%%%%%%%%%%%%%%%%%%%%%%%%%%%%%%%%%%%%%%%%%%%%%%%%%%%%%%%%
%%%%%%%%%%%%%%%%%%%%%%%%%%%%%%%%%%%%%%%%%%%%%%%%%%%%%%%%%%%%%%%%%%%%%%%%%%%%%%
\section{INTRODUCTION}\label{ch1}
Heavy-ion collisions serve as a tool to study the phase diagram of Quantum Chromodynamics (QCD), the theory that describes the strong interaction.
The QCD phase diagram is characterized by the temperature $T$ and baryon chemical potential $\mu_B$~\cite{Rajagopal:2000wf,Bzdak:2019pkr,Pandav:2022xxx,Luo:2017faz,Nonaka:2023sxt}. High $T$ and low $\mu_B$ conditions created in heavy-ion collisions likely enable the phase transition from hadronic matter to a deconfined state of quarks and gluons known as the quark-gluon plasma (QGP)~\cite{STAR:2005gfr}. Lattice Gauge Theory (LGT) calculations predict a smooth crossover transition from hadronic phase to the QGP at vanishing $\mu_B$~\cite{Aoki:2006we}. Cumulants of the net-particle number distributions have been suggested as sensitive observables in the study of QCD phase structure~\cite{Stephanov:2011pb, Friman:2011pf}. They are given by the derivative of the partition function with respect to the chemical potential and are studied extensively in theoretical frameworks like LGT~\cite{Gavai:2010zn,Gupta:2011wh,Bazavov:2020bjn,Borsanyi:2018grb} and the HRG model~\cite{Karsch:2010ck,Gupta:2022phu}, motivating their experimental investigation.

Cumulants ($C_{n}$) of event-by-event net-proton (proton $-$ antiproton) number distributions and their ratios have been measured at \sqrtsNN~=~\SI{200}{GeV} for \pp~\cite{STAR:2023zhl} and \auau~\cite{STAR:2010mib,STAR:2013gus,STAR:2020tga,STAR:2021iop,STAR:2021rls,STAR:2022vlo}
collisions at STAR. The higher-order cumulant ratios --- $C_4/C_2$, $C_5/C_1$, and $C_6/C_2$ --- show a decreasing trend from the lowest multiplicity 
% ~\footnote{Throughout this paper, the terms 'multiplicity' and 'charged-particle multiplicity' are used interchangeably, both conveying the same meaning.}
in \pp~collisions to the highest multiplicity in \auau~collisions~\cite{STAR:2023zhl}. The measurement of higher-order cumulant ratios of net-proton (taken as a proxy for net-baryon) from 0--40\% central \auau~collisions (average multiplicity $\sim$450) shows results consistent with the LGT calculations for thermalized QCD matter within uncertainties~\cite{STAR:2022vlo}. Interestingly, at the highest multiplicity ($\sim$20) from \pp~collisions the measurements also approach the LGT calculations. This observation suggests that, while the cumulant ratios systematically decrease with increasing multiplicity, the exact dependence may be collision system dependent.

In this Letter, we report the multiplicity dependence of cumulants of event-by-event net-proton number distributions and their ratios up to the sixth order in the collisions of isobaric systems --- Zirconium+Zirconium ($^{96}_{40}$Zr+$^{96}_{40}$Zr) and Ruthenium+Ruthenium ($^{96}_{44}$Ru+$^{96}_{44}$Ru) --- at \sqrtsNN~= \SI{200}{GeV}, measured by the STAR experiment. These isobar systems have the same mass number ($A=96$) but differ in proton number. We then compare the multiplicity dependence of higher-order cumulant ratios in these isobar collisions with \pp~and \auau~collisions, at the same center-of-mass energy—the highest available at RHIC for heavy-ion collisions. 
It is noteworthy that proton cumulants have also been measured at the lowest RHIC collision energy (\sqrtsNN = 3 GeV) in STAR’s Fixed Target program (FXT), where they indicate that the QCD matter created in such collisions is predominately hadronic~\cite{STAR:2021fge,STAR:2022etb,STAR:2022vlo}.
%%%%%%%%%%%%%%%%%%%%%%%%%%%%%%%%%%%%%%%%%%%%%%%%%%%%%%%%%%%%%%%%%%%%%%%%%%%%%%
%%%%%%%%%%%%%%%%%%%%%%%%%%%%%%%%%%%%%%%%%%%%%%%%%%%%%%%%%%%%%%%%%%%%%%%%%%%%%
\section{DATA ANALYSIS}\label{ch2}
Approximately 2 billion events are analyzed for each isobar collision system, selected using minimum bias triggers, which identify collisions based on signals in the trigger detectors above the noise threshold~\cite{STAR:2008med}. The main STAR sub-detectors utilized in this analysis are the Time Projection Chamber (TPC)~\cite{Anderson:2003ur} and the Time of Flight (TOF) detector~\cite{Geurts:2004fn}. The TPC measures a particle's trajectory and energy loss, which are used for obtaining the point of collision (primary vertex)
and for particle identification. The primary vertex position is required to be between \SI{-35}{cm} and \SI{25}{cm} along the beam direction and radially within \SI{2}{cm} from the center of the TPC. The selection for the primary vertex in the beam direction is asymmetric due to a timing offset in the online vertex triggering~\cite{STAR:2021mii}.   

\begin{figure*}
    \centering
     \includegraphics[width=\linewidth]{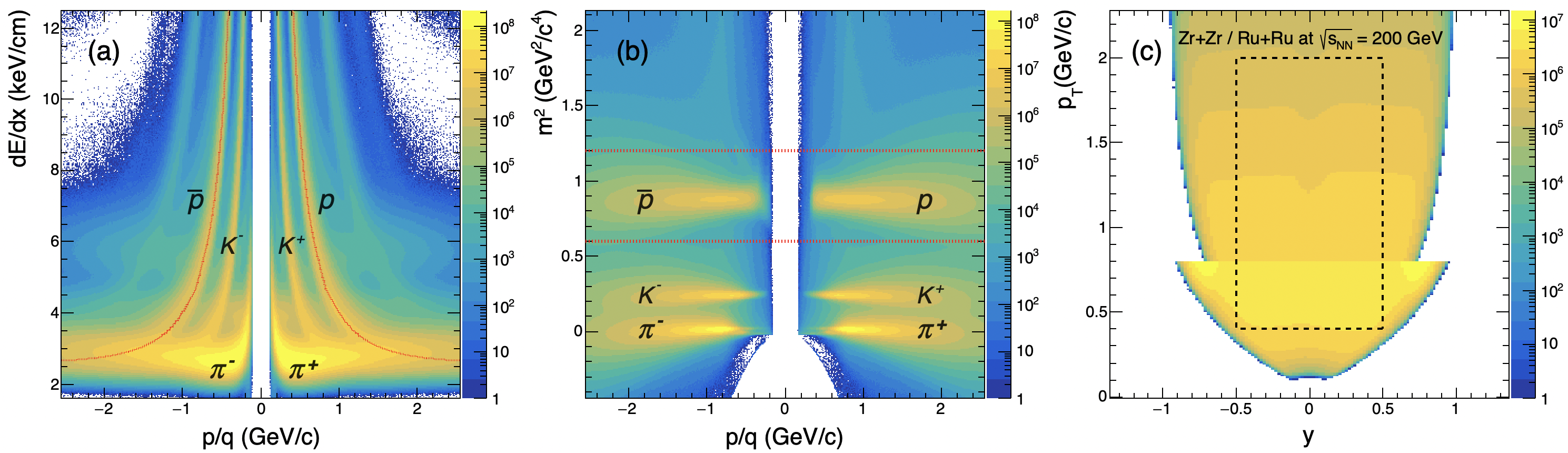}
    \caption{
    % \textcolor{blue}{(Temporary figure as an example)}  % ($\langle dE/dx \rangle$)
    Panel (a): Energy loss ($\mathrm{d}E/\mathrm{d}x$) of charged particles in the TPC as a function of rigidity (p/q). The energy loss expectation from the Bichsel function is drawn as red lines for protons and antiprotons. The letter '$p$' stands for protons, '$\bar{p}$' antiprotons, '$K^{\pm}$' for charged kaons, '$\pi^{\pm}$' for charged pions. Panel (c): Reconstructed mass squared ($m^2$) of the particle tracks in TOF as a function of rigidity. The red dashed line indicates the $m^2$ range used for the (anti-)proton selection for the analysis, i.e., $0.6 < m^2 < 1.2\;GeV^2/c^4$. Panel (c): Kinematic acceptance in transverse momentum and rapidity. The rectangular region enclosed by black dashed lines shows the acceptance of selected tracks in this analysis. The z-scale for each of these plots represents track density. }
    \label{label_fig1}
\end{figure*}

\begin{figure}
    \includegraphics[scale=0.35]{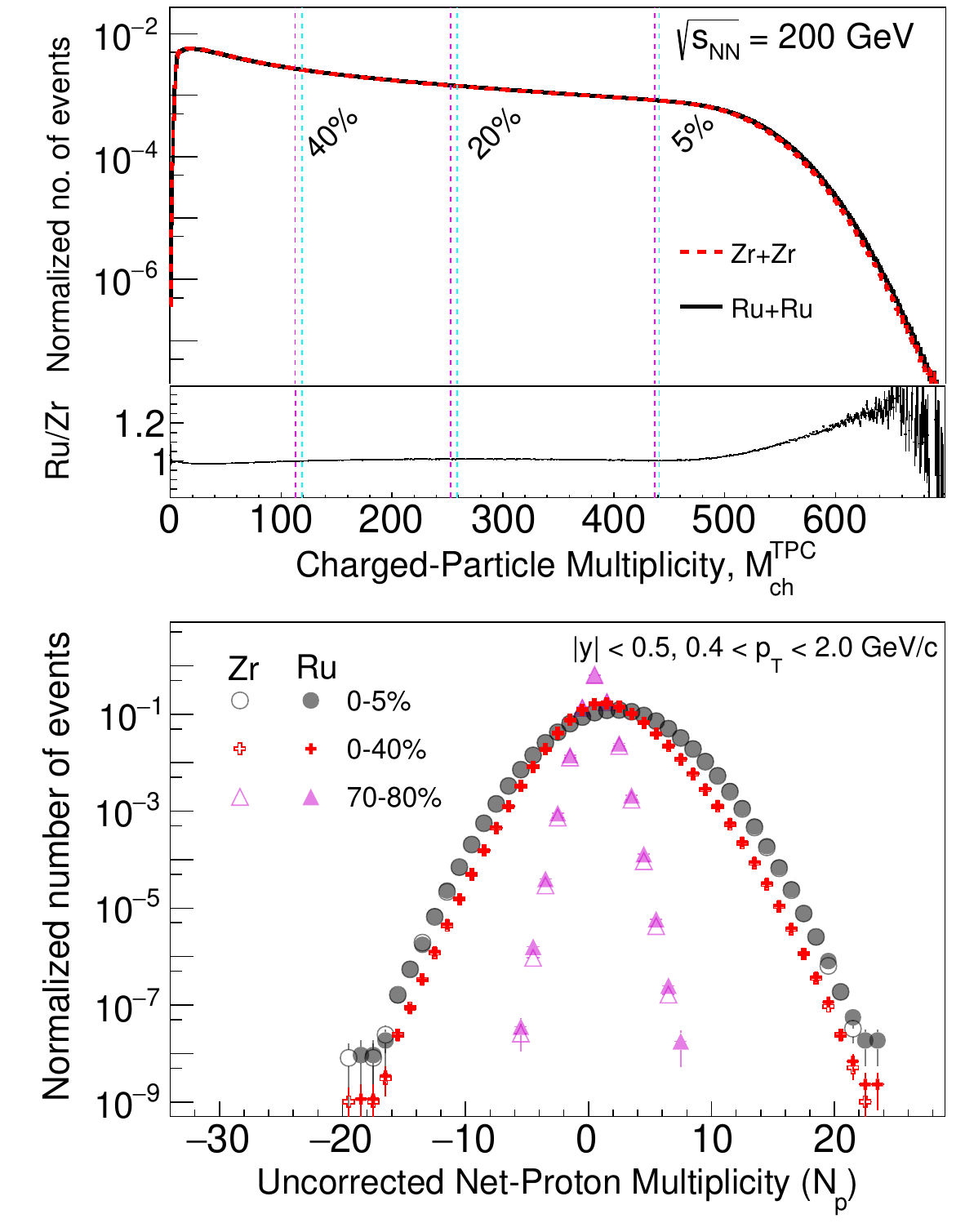}
    \caption{
    Upper plot: Charged-particle multiplicity distributions, $M_{\text{ch}}^{\text{TPC}}$, for \zrzr~and~\ruru~collisions at \sqrtsNN~=~200~\si{GeV} along with the ratio between the distributions (bottom panel). The vertical dashed magenta and cyan lines indicate the lower boundaries of the centrality classes in \zrzr~and \ruru~collisions, respectively, for the top 5\%, 20\%, and 40\% centralities.
    Lower plot: Net-proton multiplicity distributions, uncorrected for detector efficiency, in \zrzr~and~\ruru~collisions at \sqrtsNN~=~200~\si{GeV} for selected centralities. 
    }
    \label{label_fig2}
\end{figure}

Proton and antiproton tracks are selected by comparing the measured energy loss ($\mathrm{d}E/\mathrm{d}x$) in the TPC with the theoretical expectation from the Bichsel function~\cite{Bichsel:2006cs}. Panel (a) of Fig.~\ref{label_fig1} shows the measured energy loss along with the Bichsel function expectation for various charged particles, including protons. To improve particle identification at transverse momentum ($p_T$) greater than 0.8 GeV/$c$, we use mass measurements made by the TOF (Fig.~\ref{label_fig1}(b) shows (anti-)proton selection from TOF). The acceptance for (anti\mbox{-})proton is shown as a black square in Fig.~\ref{label_fig1}(c), covering rapidity $|y|<0.5$ and transverse momentum $0.4 < p_T < 2.0$ GeV/$c$. Within this acceptance, a purity of $\sim$99\% or higher is obtained for identified (anti-)protons. The distance of closest approach (DCA) of (anti-)proton tracks to the primary vertex is required to be within 1 cm to suppress contamination from secondary particles, such as those produced by the interactions with detector material~\cite{STAR:2013gus}. For good track quality, tracks having more than 20 associated TPC space points are selected~\cite{STAR:2021iop}.
%%%%%%%%===================================
%%%%%%%%===================================
% We note that, although centrality is defined using uncorrected charged-particle multiplicity, the asymptotic trends of the cumulant ratios with centrality/multiplicity—specifically, their approach toward LGT predictions and thermal expectations, which form the focus of our study—are not affected.
The centrality of collisions is characterized by the uncorrected charged-particle multiplicity measured by the TPC ($M_{\text{ch}}^{\text{TPC}}$) in the pseudo-rapidity acceptance of $|\eta|<1$ excluding (anti\mbox{-})protons to avoid self-correlation~\cite{Chatterjee:2019fey}. The resulting distribution is shown in the upper panel of Fig.~\ref{label_fig2}. A~large number of charged particles used for the centrality definition ensures good centrality resolution, effectively suppressing centrality-induced biases in cumulant measurements~\cite{Luo:2013bmi}. The ratio between the two distributions (\ruru/\zrzr) shown in the bottom panel deviate from unity, especially at the most central collisions. This is understood to be caused by the different nuclear density distributions of the colliding systems~\cite{Xu:2017zcn, Li:2018oec,Hammelmann:2019vwd}. The distributions are fitted with a Monte Carlo Glauber model simulation~\cite{Miller:2007ri} to extract geometrical information such as the average number of participating nucleons ($\langle N_{\rm part} \rangle$). By integrating the fitted distributions, the events are categorized into centrality classes. Since centrality classes are defined directly from charged-particle multiplicity, references to “multiplicity dependence” and “centrality dependence” are used interchangeably in this paper.
The centrality classes and the corresponding $\langle N_{\rm part} \rangle$ values obtained from the Glauber model fits are listed in Table~\ref{label_tab1}. The vertical dashed lines in the upper plot of Fig.~\ref{label_fig2} represent the lower edge of each centrality class.

Net-proton distributions are obtained for various centrality classes to measure their cumulants. The lower plot of Fig.~\ref{label_fig2} presents the net-proton multiplicity distributions, uncorrected for detector efficiency (i.e., the raw distributions), for the 0--5\%, 0--40\%, and 70--80\% centrality classes. Cumulants up to the sixth order are defined as the following:
%%%%%%%%===================================
\begin{equation}\label{eq:cumulants}
    \begin{aligned}
        C_1 &= \left \langle N \right \rangle  , \\
        C_2 &= \left \langle \left ( \delta N \right )^2 \right \rangle ,\\
        C_3 &= \left \langle \left ( \delta N \right )^3 \right \rangle ,\\
        &\phantom{=}\\[-\baselineskip]
        C_4 &= \left \langle \left ( \delta N \right )^4 \right \rangle
                - 3 \left \langle \left ( \delta N \right )^2 \right \rangle^2 ,\\
        C_5 &= \left \langle \left ( \delta N \right )^5 \right \rangle
                - 10 \left \langle \left ( \delta N \right )^2 \right \rangle
                        \left \langle \left ( \delta N \right )^3 \right \rangle ,\\
        C_6 &= \left \langle \left ( \delta N \right )^6 \right \rangle
                + 30 \left \langle \left ( \delta N \right )^2 \right \rangle^3 \\
            &\phantom{=} - 15 \left \langle \left ( \delta N \right )^2 \right \rangle
                    \left \langle \left ( \delta N \right )^4 \right \rangle 
                - 10 \left \langle \left ( \delta N \right )^3 \right \rangle^2. \\
        \end{aligned}
\end{equation}
%%%%%%%%===================================
The symbol $N$ represents the raw net-proton number in a given event and $\delta N$  ($\delta N=N-\langle N \rangle$) is its deviation from the average value over all events $\langle N \rangle$. The $n$-th order cumulant ($C_n$) is related to the $n$-th order susceptibility ($\chi^n$), utilized in various theoretical calculations~\cite{Karsch:2010ck,Gupta:2011wh}.
% The $n$-th order cumulants ($C_n$) are related to the $n$-th order susceptibility ($\chi^n$) calculated in various theoretical calculations~\cite{Karsch:2010ck,Gupta:2011wh} as follows: \begin{equation}\label{eq:sus}
% C_n = VT^3 \chi^n , 
% \end{equation}
% where $V$, and $T$ represent the volume and temperature of the system, respectively.
Taking the ratio of the cumulants cancels out the volume dependence and the ratios can be directly compared to theoretical calculations under the assumption of a thermalized system in a fixed volume~\cite{Karsch:2010ck, Gavai:2010zn, Gupta:2011wh, Garg:2013ata}. Cumulants and their ratios of measured net-proton distributions (e.g., those shown in Fig.~\ref{label_fig2}) are corrected for detector efficiency assuming a binomial detector response~\cite{Bzdak:2012ab,Kitazawa:2012at,Luo:2014rea,Nonaka:2017kko,Luo:2018ofd}. An event-weighted averaging, called the centrality-bin-width correction, was applied to measurements within each centrality class to account for the finite width of the class~\cite{Luo:2013bmi,STAR:2021iop,STAR:2025zdq}. The statistical uncertainties on measurements are estimated via the Delta theorem method~\cite{Luo:2011tp,Pandav:2018bdx,Pandav:2022xxx}.

To evaluate systematic uncertainties, the following quantities for (anti-)proton tracks are varied: the track quality, reflected by the number of fitting points in track reconstruction, the $\mathrm{d}E/\mathrm{d}x$ and $m^2$ cuts used for (anti-)proton identification, the DCA of the tracks, and the detector efficiency. The choices of variations considered are the same as used in STAR's prior net-proton fluctuation studies~\cite{STAR:2021iop}. All sources of systematics are subjected to a Barlow check~\cite{Barlow:2002yb} and those whose variations exceed statistical expectations are then added in quadrature to obtain the total systematic uncertainties. As an example, the magnitude of systematic uncertainty on net-proton $C_4/C_2$ ($C_6/C_2$) from 0--40\% Zr+Zr collisions is found to be 0.014 (0.189). For Ru+Ru collisions, the corresponding value is 0.015 (0.281) for $C_4/C_2$ ($C_6/C_2$) measured in 0--40\% centrality. In general, the magnitude of systematic uncertainties between the two systems is comparable.
%%%%%%%%===================================
%%%%%%%%===================================
%%%%%%%%===================================
\begin{table}[h!]
\caption{
The average number of participant nucleons ($\langle N_{\rm part} \rangle$) along with corresponding systematic uncertainty estimates for centrality classes from Glauber model fit to experimental data in \zrzr~and~\ruru~collisions. 
}
\begin{tabular}{c|cc}
\toprule
 & \zrzr & \ruru \\ 

Centrality (\%) & \multicolumn{2}{c}{$\langle N_{\rm part} \rangle$} \\ 
\hline
0--5 & $166 \pm 1$ & $167 \pm 1$ \\
5--10 & $144 \pm 3$ & $144 \pm 3$ \\
0--20 & $134 \pm 3$ & $135 \pm 3$ \\
10--20 & $113 \pm 4$ & $115 \pm 4$ \\
20--30 & $81 \pm 5$ & $83 \pm 5$ \\
0--40 & $102 \pm 4$ & $103 \pm 4$ \\
30--40 & $57 \pm 5$ & $59 \pm 5$ \\
40--50 & $38 \pm 5$ & $40 \pm 5$ \\
50--60 & $25 \pm 5$ & $26 \pm 5$ \\
60--70 & $15 \pm 4$ & $16 \pm 4$ \\
70--80 & $9 \pm 3$ & $9 \pm 3$ \\

\end{tabular}
\label{label_tab1}
\end{table}
%%%%%%%%===================================

%%%%%%%%%%%%%%%%%%%%%%%%%%%%%%%%%%%%%%%%%%%%%%%%%%%%%%%%%%%%%%%%%%%%%%%%%%%%%%
%%%%%%%%%%%%%%%%%%%%%%%%%%%%%%%%%%%%%%%%%%%%%%%%%%%%%%%%%%%%%%%%%%%%%%%%%%%%%%
\section{RESULTS AND DISCUSSIONS}\label{ch3}
\begin{figure*}[!htbp]
\centering
\includegraphics[width=\linewidth]{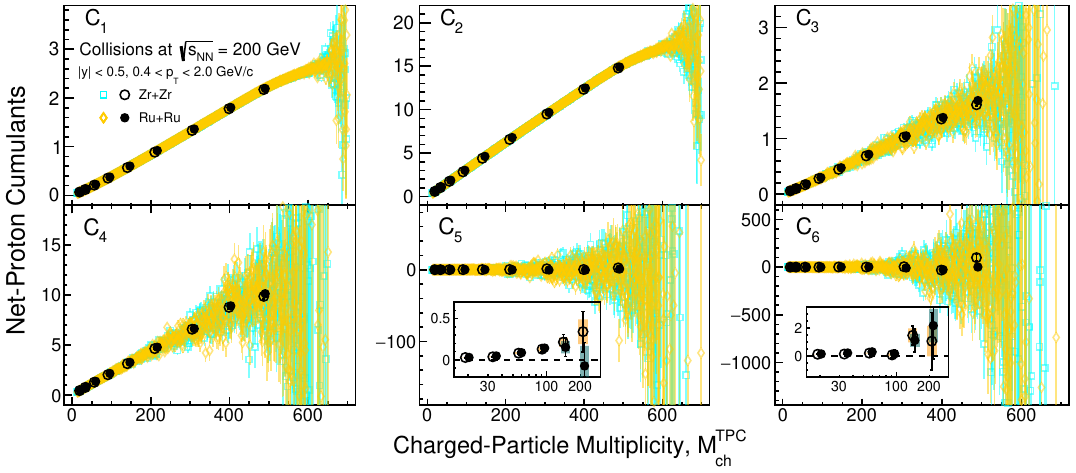}
\caption{Net-proton cumulants from the first ($C_1$) to the sixth order ($C_6$) as a function of charged-particle multiplicity, $M_{\text{ch}}^{\text{TPC}}$, for \zrzr~(cyan squares and open circles) and~\ruru~(orange diamonds and filled circles) collisions at \sqrtsNN~=~200~\si{GeV}. Circle markers represent centrality-bin-width averaged cumulants for nine centrality classes from 0--5\%, 5--10\%, 10--20\%, 20--30\%, and so on, up to 70--80\%. Each point is plotted with the corresponding statistical uncertainties. The inset panels present $C_5$ and $C_6$ results from 70--80\% to 20--30\% centrality. Systematic uncertainties are shown as colored bands on the data points (orange for \zrzr~and blue for \ruru~collisions).}
\label{label_fig3}
\end{figure*}

Figure~\ref{label_fig3} shows the net-proton cumulants in isobar collisions from the first to the sixth order. The cumulants are presented for unit charged-particle multiplicity bins as well as for various centrality classes where they have been centrality-bin-width averaged. The cumulants $C_5$ and $C_6$ from 70--80\% to 20--30\% centrality are shown in zoomed insets. Multiplicity-driven growth is observed for all cumulants, albeit with larger uncertainties at higher-orders. The slopes obtained from a linear fit are positive beyond $5\sigma$ significance for both systems except for $C_6$ where the significance level is $\sim$1$\sigma$ for \zrzr~ and $\sim$3$\sigma$ for \ruru.
%%%%%%%%===================================
%%%%%%%%===================================
%%%%%%%%===================================

\begin{figure*}
    \centering
    % \vstretch{1}    {\includegraphics[width=1.00\columnwidth,trim={0 0 0 0 },clip]{Figures_/Isobar_Fig4_26Aug2025.pdf}}   
\includegraphics[scale=0.27]{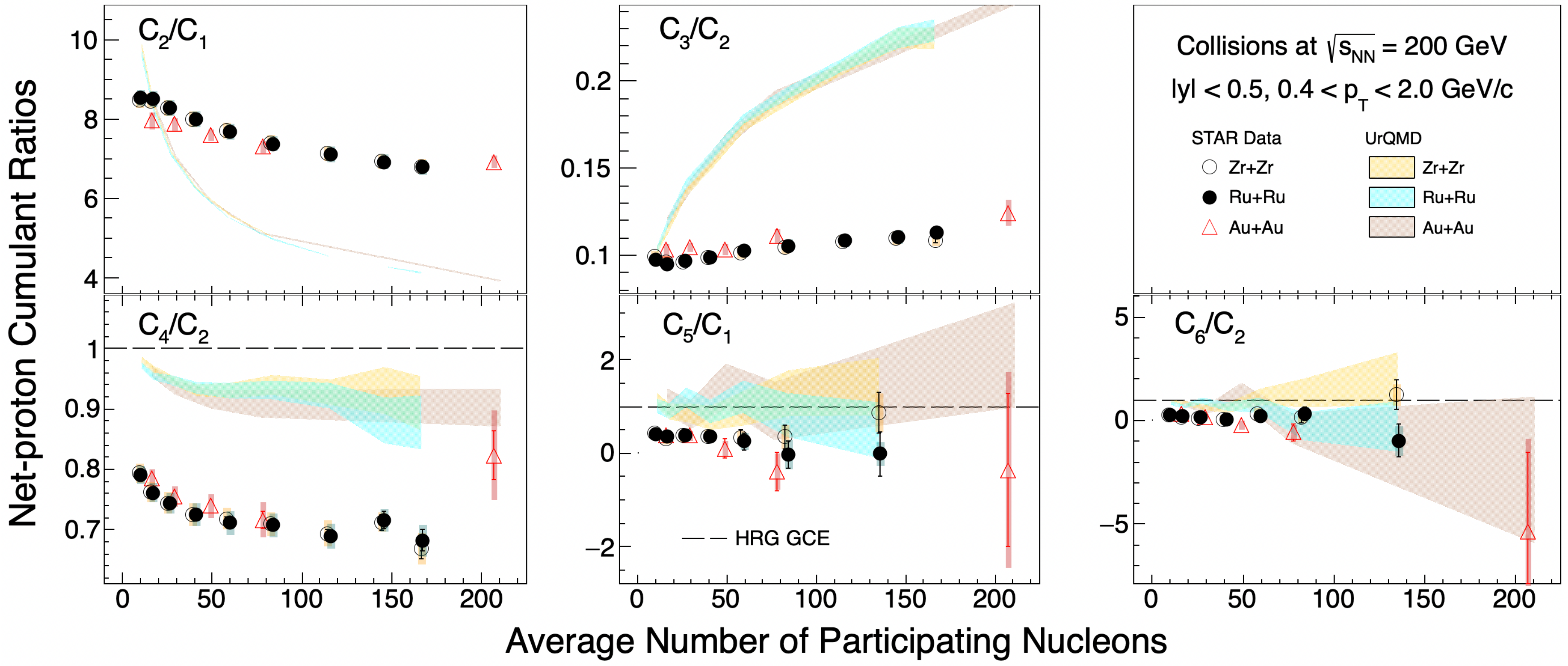}
    \caption{
    The collision centrality dependence (shown as a function of average number of participating nucleons) of the net-proton cumulant ratios for \zrzr~(open circles), \ruru~(filled circles) at \sqrtsNN~=~200~\si{GeV}. Results from~\auau ~collisions~\cite{STAR:2021iop, STAR:2021rls, STAR:2022vlo}~(open triangles) at the same collision energy are also shown for comparison. Bars and colored bands on the data points represent statistical and systematic uncertainties, respectively.
    For \auau~collisions~\cite{STAR:2021iop, STAR:2021rls, STAR:2022vlo}, the most central point corresponds to the 0--40\% centrality class.
    In the case of the $C_5/C_1$ and $C_6/C_2$ from the isobar collisions, the most central point corresponds to the 0--20\% centrality class. Expectations from the UrQMD model, including statistical uncertainties, are shown as bands. The dashed line represents the calculations from the thermal model HRG in a grand canonical ensemble (HRG GCE). By definition, the HRG GCE expectation for $C_4/C_2$, $C_5/C_1$, and $C_6/C_2$ is unity. 
    }
    \label{label_fig4}
\end{figure*}

Figure~\ref{label_fig4} shows the net-proton cumulant ratios: $C_2/C_1$, $C_3/C_2$, $C_4/C_2$, $C_5/C_1$ and $C_6/C_2$ from \zrzr~and Ru+Ru collisions as a function of collision centrality (expressed as average number of participating nucleons, $\langle N_{\rm part} \rangle$). These ratios have been widely studied by experiments at various collision energies and have been tested against various theoretical models~\cite{STAR:2020tga,STAR:2021iop,STAR:2021rls,STAR:2022vlo,STAR:2023zhl,STAR:2021fge,STAR:2022etb,ALICE:2022xpf,HADES:2020wpc}. In Fig.~\ref{label_fig4}, we observe a smooth variation of cumulant ratios with respect to centrality and the results from the two isobar systems are consistent within uncertainties. The results from \auau~collisions~\cite{STAR:2021iop, STAR:2021rls, STAR:2022vlo} are also shown for comparison. The \auau~data seem to align with the observed centrality dependence in isobar collisions within uncertainties except for $C_2/C_1$ and $C_3/C_2$ in peripheral collisions where some differences ($\lesssim$6\%) can be seen. The thermal HRG model calculations in the Grand Canonical Ensemble (HRG GCE, shown for $C_4/C_2$, $C_5/C_1$, and $C_6/C_2$) are at unity by definition and cannot explain the observed centrality dependence in the experimental data. Calculations from the hadronic transport model UrQMD~\cite{Bleicher:1999xi,Bass:1998ca} also fail to adequately explain the observed trends for all three systems. For the lowest two cumulant ratios, the model results misses the measured trends completely. For the higher-order ratios, the UrQMD results remain largely flat and deviate from the data, except as one moves into the more central bins of these systems, where the large uncertainties in both data and model render them compatible. The centrality definition in UrQMD model calculations is performed in the same way as for the experimental data.  
%%%%%%%%===================================
%%%%%%%%===================================

\begin{figure*}
    \centering
     \includegraphics[width=\linewidth]{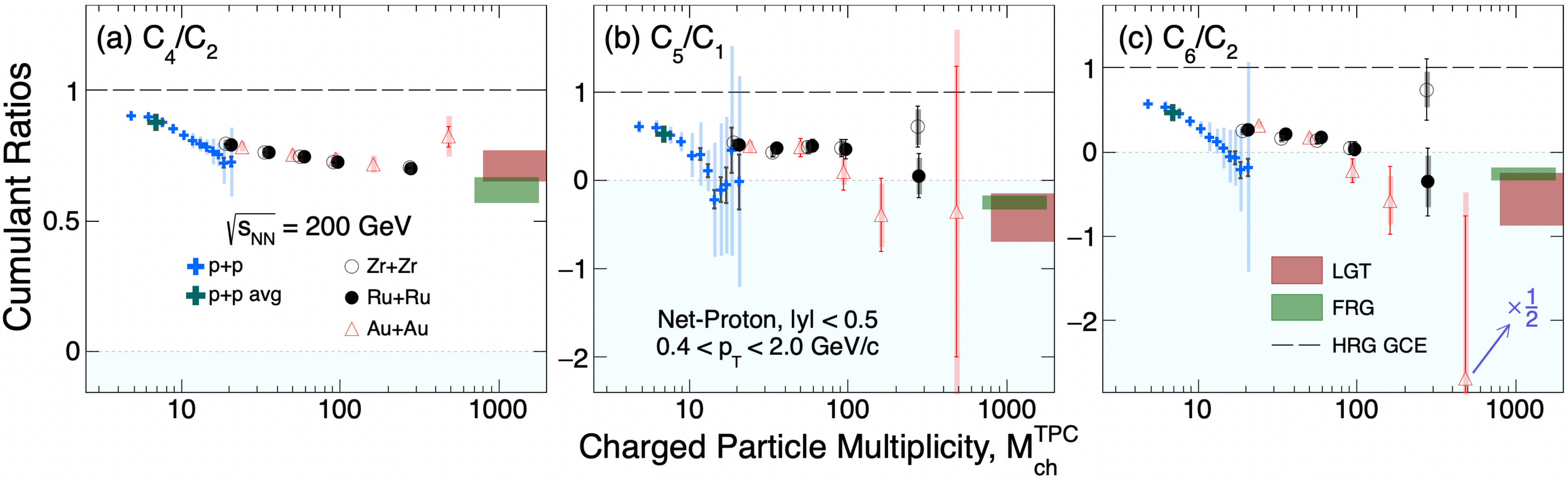}
    \caption{
    Net-proton cumulant ratios $C_4/C_2$, $C_5/C_1$, and $C_6/C_2$ from \pp~\cite{STAR:2023zhl}\,(blue), \zrzr~and \ruru\,(black), and \auau~\cite{STAR:2021iop, STAR:2021rls, STAR:2022vlo}\,(red markers) collisions at \sqrtsNN~=~200~\si{GeV} shown as a function of measured charged-particle multiplicity, $M_{\text{ch}}^{\text{TPC}}$. 
    % The efficiencies of the charged-particle multiplicity are corrected to match that of the \auau~collisions.
    For better statistical precision, results from a wide 0--\SI{40}{\%} centrality are shown as the top centrality for the isobar and \auau~collisions, along with those from 40--50\%, 50--60\%, 60--70\% and 70--80\% centrality classes. Model calculations from Lattice Gauge Theory (LGT)~\cite{Bazavov:2020bjn} (brown bands), FRG~\cite{Fu:2021oaw} (green bands), and HRG GCE (dashed lines at unity) are also shown. The $C_6/C_2$ (0-40\%) data for Au+Au has been scaled down by factor of 0.5 for clarity of presentation. 
    }
    \label{label_fig5}
\end{figure*}

Figure~\ref{label_fig5} shows the higher-order cumulant ratios $C_4/C_2$, $C_5/C_1$, and $C_6/C_2$ of net-proton distributions from different colliding systems -- \pp~\cite{STAR:2023zhl}, isobars, and \auau~\cite{STAR:2021rls,STAR:2022vlo} at \sqrtsNN~=~\SI{200}{GeV} as a function of charged particle multiplicity. For better statistical precision, the top centrality class shown for the isobar data corresponds to a wide 0\mbox{-}40\% centrality bin. \auau~data are shown for comparison using the same centrality binning. From \pp~to isobar to \auau, the presented cumulant ratios, in general, decrease with increasing multiplicity. Mild deviations from this trend can be seen for $C_4/C_2$ in Au+Au and $C_5/C_1$ and $C_6/C_2$ in Zr+Zr collisions for 0--40\% centrality, but they remain comparable within the range of uncertainties. The measured net-proton cumulant ratios are compared with theoretical calculations from LGT~\cite{Bazavov:2020bjn} and QCD-based functional renormalization group model (FRG)~\cite{Fu:2021oaw}, bearing in mind caveats discussed in Ref.~\cite{STAR:2021rls}. Both calculations include the formation of thermalized QCD matter via a smooth crossover transition and predict a negative value for net-baryon $C_5/C_1$ and $C_6/C_2$. The \pp~and \auau~measurements at high multiplicity, as reported earlier~\cite{STAR:2023zhl,STAR:2021rls,STAR:2022vlo}, showed a systematic trend toward negative $C_5/C_1$ and $C_6/C_2$, in agreement with theoretical expectations within uncertainties. The new results from \ruru~collisions are observed to approach negative values at the highest multiplicities, trending toward the LGT and FRG calculations. The $C_5/C_1$ data, however, has a weak dependence on multiplicity. The data from \zrzr~collisions show a similar trend and closely follow those from \ruru~as a function of multiplicity, with only slight deviations ($\lesssim 1.7\sigma$) in $C_5/C_1$ and $C_6/C_2$ at the highest multiplicity. The $C_4/C_2$ from the two isobaric systems also converges towards the LGT and FRG calculations at the highest multiplicity. The HRG GCE fails to explain the observed trends in the experimental data.

%%%%%%%%===================================
%%%%%%%%===================================
%%%%%%%%%%%%%%%%%%%%%%%%%%%%%%%%%%%%%%%%%%%%%%%%%%%%%%%%%%%%%%%%%%%%%%%%%%%%%%
%%%%%%%%%%%%%%%%%%%%%%%%%%%%%%%%%%%%%%%%%%%%%%%%%%%%%%%%%%%%%%%%%%%%%%%%%%%%%%
\section{SUMMARY}\label{ch4}
We report the new STAR measurements on cumulants of event-by-event net-proton distributions and their ratios up to the sixth order as a function of charged-particle multiplicity and collision centrality in \zrzr~and~\ruru~collisions at \sqrtsNN~=~\SI{200}{GeV}. All orders of cumulant are found to increase with centrality. A smooth variation for cumulant ratios with respect to centrality is observed and the results from the two isobar systems are consistent within uncertainties. The transport model UrQMD and the thermal model HRG GCE  fail to explain the observed centrality dependence. A comparison is made with data from \auau~and \pp~collision systems at the same center-of-mass energy. In general, the higher-order cumulant ratios $C_4/C_2$, $C_5/C_1$, and $C_6/C_2$ across systems show a progressively decreasing trend with increasing multiplicity. The observed trends in the isobar data fit well with those from Au+Au collisions, within uncertainties. As proxies for net-baryon cumulant ratios, the net-proton $C_4/C_2$, $C_5/C_1$, and $C_6/C_2$ are compared to LGT calculations. The higher-order cumulant ratios, within uncertainties, exhibit an overall tendency to approach LGT predictions with increasing multiplicity. This could imply that the medium created in these heavy-ion collisions gradually evolves into thermalized QCD matter, undergoing a crossover transition with multiplicity. Precision measurements from different collision systems and energies will provide a more systematic understanding of the thermalization of matter created in high-energy collisions. STAR's ongoing high-statistics \auau~ run at \sqrtsNN~=~\SI{200}{GeV}, benefitted by the upgraded inner TPC and thus wider kinematic acceptance, will significantly aid this effort. 

We thank the RHIC Operations Group and SDCC at BNL, the NERSC Center at LBNL, and the Open Science Grid consortium for providing resources and support. This work was supported in part by the Office of Nuclear Physics within the U.S. DOE Office of Science, the U.S. National Science Foundation, National Natural Science Foundation of China, Chinese Academy of Science, the Ministry of Science and Technology of China and the Chinese Ministry of Education, NSTC Taipei, the National Research Foundation of Korea, Czech Science Foundation and Ministry of Education, Youth and Sports of the Czech Republic, Hungarian National Research, Development and Innovation Office, New National Excellency Programme of the Hungarian Ministry of Human Capacities, Department of Atomic Energy and Department of Science and Technology of the Government of India, the National Science Centre and WUT ID-UB of Poland, the Ministry of Science, Education and Sports of the Republic of Croatia, German Bundesministerium f\"ur Bildung, Wissenschaft, Forschung and Technologie (BMBF), Helmholtz Association, Ministry of Education, Culture, Sports, Science, and Technology (MEXT), Japan Society for the Promotion of Science (JSPS) and Agencia Nacional de Investigaci\'on y Desarrollo (ANID) of Chile.
%%%%%%====================================
%%%%%%====================================

\bibliographystyle{apsrev4-1}
\bibliography{References_Isobar.bib}

\end{document}